# Monte Carlo Simulation of Spallation and Fission Fragment Distributions for ADS-Related Nuclear Reactions

SunWenming[1]

1(Graduate School of Science, The University ofTokyo, Tokyo, 113-0033)

**Monte Carlo simulations with the CRISP code were conducted to study spallation and fission fragment distributions induced by intermediate- and high-energy protons and photons on actinide and pre-actinide nuclei. The model accounts for intranuclear cascade, pre-equilibrium, and evaporation-fission competition, enabling consistent treatment of both residues and fission products. Comparisons with experimental data show good agreement in mass and charge distributions, with minor deviations for light fragments. The results highlight the reliability of Monte Carlo approaches for predicting residual nuclei and fragment yields under accelerator-driven system (ADS) conditions. This work provides nuclear data relevant to ADS design, safety, and transmutation analysis**

## 1. Introduction

The accelerator driven system (ADS) is an innovative reactor which is being developed as a dedicated burner in a double strata fuel Cycle to incinerate nuclear waste [1–4]. The ADS system consists of a subcritical assembly driven by accelerator delivering a proton beam on a target to produce neutrons by a spallation reaction. The spallation target constitutes at the same time the physical and the functional interface between the accelerator and the subcritical reactor. For this reason it is probably the most innovative component of the ADS, and its design is a key issue to develop ADS. The performance of the reactor is characterized by the number of neutrons emitted for incident proton, the mean energy deposited in the target for neutron produced, the neutron spectrum, and the spallation product distribution [5].

The detailed design of spallation neutron sources or accelerator-driven systems (ADS) requires reliable computational tools in order to optimize their performance in terms of useful neutron production and to properly assess specific problems likely to happen in such systems. Among those problems are the radioactivity induced by spallation reaction and the problem of shielding [6] the radiation due to energetic particles generated in the reaction, the changes in the chemical composition and radiation damage in target, window, or structure materials [7], and induced radiotoxicity within the target [8] due the production of several nuclides. Radiation damage can arise from gas production that causes embrittlement of structural materials and from atomic displacements (DPA) which fragilize the various components of the spallation source. Modifications of the chemical composition of these materials possibly result into problems of corrosion or alloy cohesion and modification of mechanical properties because of the appearance of compounds not existing initially in the materials.

It is important to emphasize that, at the present status in the development of ADS, it is still necessary to study the best technological options for variables such as the material used for the target and the energy of incident particles. For this reason the most important aspect in the calculations of yields of nuclides is the confidence one has on the results obtained. More important than accuracy, at this moment, may be the reliability of the calculation method. One should consider more important methods which take into consideration the correct mechanisms of reactions, than those methods that have many free parameters to fit experimental data. These are of great importance to interpolate experimental data, when they are available, but are of restrict use when there is not any data.

This paper is organized as follows: Section 2 describes the main mechanisms for fragments production in nuclear reactions and gives a short description of the intranuclear cascade process, which is relevant for the description of the residual nucleus formation. In Section 3 the evaporation mechanism is described. This is the most important mechanism for the study of spallation. In Section 4 the fission process is described, this decay channel being relevant for the formation of fission fragments. In Section 5 we present and discuss the results obtained with the CRISP code, and in Section 6 we show our conclusions.

## 2. Production of Nuclide in Nuclear Reactions

The production of nuclide is associated to the different mechanisms through which the nuclear reaction evolves. Depending on the target, on the probe and on the energy, different mechanisms are more or less relevant. On the other hand, the distribution of fragments generated when the reaction finishes is strongly dependent on the channels, open to the system and on the respective branching ratios.

As the reaction energy increases, more channels are available and the complexity of the reaction increases. This high complexity is the main reason to adopt Monte Carlo (MC) methods in calculations of nuclear reactions results. Indeed the large number of particles, the large number of available channels and the fact that the reaction may be understood as a Markovian sequence of steps are features that perfectly match the characteristics of Monte Carlo calculations. In this work we adopt the MC method and use the implementation for nuclear reaction calculations developed in the CRISP code [12], which has been already used in ADS studies [13–15]. This implementation is described in some details below.

It is well established that intermediate- and high-energy reactions follow a two-step mechanism, as proposed by Bohr. In the first one, usually called intranuclear cascade, the energy and momentum of the incident particle is distributed among a few nucleons through baryon-baryon reactions which are mainly elastic, but at sufficiently high energies nucleonic degrees of freedom can be excited. This step finishes when there is not any nucleons with kinetic energy that is high enough to escape from the nuclear binding. Thereafter the collisions among nucleons lead only to the system thermalization.

The second step depends on the excitation of the residual nucleus formed at the end of the intranuclear cascade. If the nuclear excitation energy is $E/A \leq 3$ MeV, a statistical competition between evaporation and fission takes place. For heavy-nuclei ($A > 230$) fission is the dominant channel, and in most of the cases the reaction ends with the formation of two fragments [16]. For nuclei with $A < 230$, fission is much less probable, and in most cases the nucleus evaporates until there is not energy for evaporation of any particle (neutrons, protons, and alpha particles are the most frequent). Then a spallation product is formed which is characterized by its mass and atomic numbers.

If the nuclear excitation energy is $E \geq 3$MeV/$A$ an entirely different process may occur, namely, the nuclear multifragmentation [17–19]. This process is much faster than the evaporation/fission competition, and it is characterized by the simultaneous production of a large number of fragments with very different mass and atomic numbers. In the following we give a brief description of each process mentioned above.

*2.1. Intranuclear Cascade.* We now describe the nuclear mechanisms relevant in nuclear reactions at intermediate and high energies, which have been implemented in the CRISP code. The first thing we have to deal with is the intranuclear cascade (IC). As the incident particle enters the nucleus region, IC is the first mechanism of the reaction which is triggered by the collision of the probe with one of the nucleons in inelastic scattering. This collision, called elementary collision, will always generate secondary particles, which can be two nucleons in the case of elastic scattering, mesons, or resonance states of the nucleon. These secondaries have energies relatively high compared to the other nucleons in the Fermi sea and occupy high-energy single-state levels in the system [20–23]. We call them cascade particles.

The secondary particles generated in an elementary collision will propagate inside the nucleus and may interact with other particles, increasing the number of cascade particles or may reach the nuclear surface. In this case, if it has energy higher than the nuclear binding energy, it escapes from the nucleus, otherwise it is reflected back and keeps its propagation inside the nucleus. In this way, as the intranuclear cascade continues, the number of cascade particles increases. The decision to stop the intranuclear cascade and start the second step of the reaction is based on energetic criteria, namely, when there is not any bound particle that is in an excited state or has kinetic energy greater than its binging energy [21, 22, 24].

There are several important aspects in the intranuclear cascade calculations with the CRISP code. The first one is that it is a multicollisional simulation of the cascade, with all nucleons moving simultaneously. The time-ordered sequence of elementary collisions considers the probability of interactions among all those particles based on their respective cross sections. This is an important difference with respect to other Monte Carlo codes where only one cascade particle is considered, while all others are kept frozen in their initial state.

The multicollisional approach is an important improvement in the simulation of intranuclear cascade in respect to the MC method used in well-known codes like those from Barashenkov et al. or Bertini et al. [25, 26], since dynamical aspects as nuclear density modifications or the evolution of occupancy numbers during the cascade are naturally taken into account in the former method but not in the last [24].

Another important aspect is the Pauli blocking mechanism, responsible for keeping track of possible violations of the Pauli Principle. With the multicollisional method it is possible to adopt a very precise method for verifying the availability of single-particle states for the fermions generated in the elementary collisions, eliminating possible violations of this important principle of quantum mechanics. It is because we use this method for Pauli blocking that we can use the energetic criteria to decide to end the intranuclear cascade [27].

After the cascade finishes there is not any particle with energy high enough to escape from the nucleus. Then a sequence of elementary collisions will distribute the excitation energy remaining in the nucleus among all nucleons. This is called thermalization process. The main characteristics of the nucleus do not change during this step, and the mass number, the atomic number, and the excitation energy at the end of the thermalization step are the same as in the end of the intranuclear cascade.

In a CRISP simulation, the reactions can be initiated by intermediate- and high-energy protons [21] or photons [21, 22, 27, 28]. It has been used for energies up to 3.5 GeV [24], where it was shown that CRISP code can give good results for total photonuclear absorption cross sections from approximately 50 MeV, where the quasi-deuteron absorption mechanism is dominant, up to 3.5 GeV, where the so-called photon-hadronization mechanism is dominant, leading to a shadowing efect in the cross-section. Recently the CRISP code has been used to study final state interactions for the nonmesonic weak decay of hypernucleus [29]. The results show that it can be used even for light nuclei, as $^{12}$C, and relatively low energies, as in the case of hypernucleus decay.

## 3. Evaporation

The thermalization is followed by the evaporation process, when nucleons or small clusters are emitted carrying part of the nuclear excitation energy. This process continues while there is energy enough in the nucleus to allow the evaporation of any particle. It consists of a sequence of emissions of particles by a nucleus, each one governed by the Weisskopf theory. Here the evaporation regime is governed by the relative probabilities of diferent particle emission channels [27, 28, 30].

These probabilities are obtained from the particle emission width, Gamma$_k$, calculated according to the well-known Weisskopf evaporation model [31] in such a way that for proton emission we have

$$\frac{\Gamma_p}{\Gamma_n} = \frac{E_p}{E_n} \exp\left\{2\left[(a_p E_p)^{1/2} - (a_n E_n)^{1/2}\right]\right\}, \quad (1)$$

and for alpha particle emission,

$$\frac{\Gamma_\alpha}{\Gamma_n} = \frac{2E_\alpha}{E_n} \exp\left\{2\left[(a_\alpha E_\alpha)^{1/2} - (a_n E_n)^{1/2}\right]\right\}, \quad (2)$$

Table 1: Values for the relevant parameters in semiempirical mass formula.

| Parameter | Value |
|---|---|
| $a_v$ | $-15.0175 \pm 0.000013$ |
| $a_{sf}$ | $15.5981 \pm 0.000032$ |
| $a_{sym}$ | $-7.09740 \pm 0.00067$ |
| $a_{ss}$ | $144.764 \pm 0.0022$ |

Table 2: Values for the relevant parameters in Dostrovsky's empirical formulas.

| Parameter | Value |
|---|---|
| $a_1$ | $18.81302 \pm 0.000097$ |
| $a_2$ | $1.30001 \pm 0.000097$ |
| $a_3$ | $18.670295 \pm 0.000097$ |
| $a_4$ | $4.23501 \pm 0.000097$ |
| $a_5$ | $18.89 \pm 0.19$ |
| $a_6$ | $24.82 \pm 2.18$ |

where $E_k$ corresponds to the nuclear excitation energy after the emission of a particle of kind $k$, with $k = p, n, \alpha$, which are calculated by

$$\begin{aligned} E_n &= E - B_n, \\ E_p &= E - B_p - V_p, \\ E_\alpha &= E - B_\alpha - V_\alpha, \end{aligned} \quad (3)$$

where $B_n$, $B_p$, and $B_\alpha$ are the separation energy of neutrons, protons, and alpha particles, and $V_p$ and $V_\alpha$ are the Coulomb potentials for protons and alpha particles, respectively.

These Coulomb potentials are given by the expressions:

$$\begin{aligned} V_p &= C \frac{K_p (Z-1) e^2}{r_0 (A-1)^{1/3} + R_p}, \\ V_\alpha &= C \frac{2 K_\alpha (Z-2) e^2}{r_0 (A-4)^{1/3} + R_\alpha}, \end{aligned} \quad (4)$$

where $K_p = 0.70$ and $K_\alpha = 0.83$ are the Coulomb barrier penetrabilities for protons and alpha particles. Also $R_p = 1.14$ fm is the proton radius, $R_\alpha = 2.16$ fm is the alpha particle radius, and $r_0 = 1.18$ fm. $C$ is the charged particle Coulomb barrier correction and is calculated according to

$$C = 1 - \frac{E}{B}, \quad (5)$$

$B$ being the nuclear binding energy.

The level density parameters $a_k$ are calculated from the Dostrovsky's empirical formulas [32]:

$$\begin{aligned} a_n &= \frac{A}{a_1}\left(1 - a_2 \frac{A-2Z}{A^2}\right)^2, \\ a_p &= \frac{A}{a_3}\left(1 + a_4 \frac{A-2Z}{A^2}\right)^2, \\ a_\alpha &= \frac{A}{a_5}\left(1 - \frac{a_6}{Z}\right)^2. \end{aligned} \quad (6)$$

Table 3: Values for some of the relevant parameters in the multimode formula for the fission-fragment mass distributions.

| Parameter | Low-energy | Parameter | $^{241}$Am | $^{237}$Np | $^{238}$U | $^{208}$Pb |
|---|---|---|---|---|---|---|
| $\Gamma_S$ | 10.0 ± 2 | $K_S$ (mb) | 2970.0 ± 20.5 | 2590.0 ± 23.3 | — | — |
| $A_1^H$ | 135.0 ± 1 | $K_1$ (mb) | 45.8 ± 0.2 | 49.0 ± 0.3 | — | — |
| $\Gamma_1^H$ | 3.75 ± 2 | $K_2$ (mb) | 220.5 ± 1.5 | 252.0 ± 1.3 | — | — |
| $A_2^H$ | 141.0 ± 2 | $\mu_1$ | 4.1 | 4.1 | 4.1 | 0.97 |
| $\Gamma_2^H$ | 5.0 ± 1 | $\mu_2$ | 0.38 | 0.38 | 0.38 | 0.413 |
| | | $\gamma_1$ | 0.92 | 0.92 | 0.92 | 0.5 |
| | | $\gamma_2$ | 0.003 | 0.003 | 0.003 | 0.008 |

Table 4: Values of the relevant parameters found for the best agreement between simulated and experimental data for the fission of $^{208}$Pb induced by 500 MeV protons. Errors indicated represent superior limit for uncertainties.

| Parameter | Value |
|---|---|
| $b_1$ | 0.01 ± 0.05 |
| $b_2$ | 121.68 ± 0.05 |
| $b_3$ | 0.23 ± 0.05 |
| $b_4$ | 125.66 ± 0.05 |
| $b_5$ | 14.93 ± 0.05 |
| $a_6$ | 3.97 ± 0.05 |
| $b_7$ | 5.21 ± 0.05 |

For the mass formula it was considered the semiempirical one proposed in [33] for the nuclear binding energy so that the nuclear masses are calculated according to

$$M(A,Z) = Zm_p + Nm_n + a_v A + a_{sf} A^{2/3} + \frac{3e^2}{5r_0} \frac{Z^2}{A^{1/3}} + \left(a_{sym} + a_{ss} A^{-1/3}\right) \frac{(N-Z)^2}{A}, \quad (7)$$

$N$ being the number of neutrons. Formula (7) was fitted to compilation of Audi et al. [34] by the method of least squares with the MINUIT package [35]. The values for each parameter corresponding to the best fit are shown in Table 1.

To evaluate the evaporation probability one assumes that it is proportional to the corresponding width, that is, [15, 30]

$$P_k = \frac{\Gamma_k}{\Gamma_n + \Gamma_p + \Gamma_\alpha}. \quad (8)$$

While enough energy is available for particle evaporation others emissions are processed. The evaporation phase ends when the excitation energy of the nucleus is smaller than all the separation energies $B_n$, $B_p$, and $B_\alpha$. At this point a nucleus that can be completely different from the initial one is formed. This nucleus is called spallation product.

## 4. Fission

The CRISP code can also evaluate the fission probability [36]. Fission is a process that competes with evaporation in the sense that each nucleus in the evaporation sequence can undergo fission, forming two fragments with masses around one-half of the fissioning nucleus. This process can be easily included in the frame of the evaporation process by including the fission branching ratio, $\Gamma_f$, in formula (8), in such a way that it now reads

$$P_k = \frac{\Gamma_k}{\Gamma_n + \Gamma_p + \Gamma_\alpha + \Gamma_f}. \quad (9)$$

The fission branching ratio is calculated according to the Bohr-Wheeler fission model [37], by

$$\frac{\Gamma_f}{\Gamma_n} = K_f \exp\left\{2\left[\left(a_f E_f\right)^{1/2} - (a_n E_n)^{1/2}\right]\right\}, \quad (10)$$

where,

$$K_f = K_0 a_n \frac{\left[2\left(a_f E_f\right)^{1/2} - 1\right]}{\left(4A^{2/3} a_f E_n\right)},$$

$$E_f = E - B^f,$$
$$a_f = r_f a_n, \quad (11)$$

$B_f$ being the fission barrier calculated according to the Nix model [38], $K_0 = 14.39$ MeV and $r_f$ an adjustable parameter.

With formula (9) it is now possible to calculate the probability of fission, $P_f$, at each step of the evaporation/fission competition process. Whenever the fission channel is chosen, two fragments are formed [39, 40], the heaviest one having mass and atomic numbers, $A^H$ and $Z^H$, respectively, is sorted according to a probability distribution given by the statistical scission model (SSM) from Brosa et al. [41]. The lighter fragment has mass and atomic numbers given, respectively, by $A^L = A^F - A^H$ and $Z^L = Z^F - Z^H$.

The Brosa's model takes into account the collective effects of nuclear deformation during fission through a liquid-drop model and includes single-particle effects through microscopic shell-model corrections. The microscopic corrections create valleys in the space of elongation and mass number, each valley corresponding to one different fission mode. Fission cross section results from the incoherent sum of

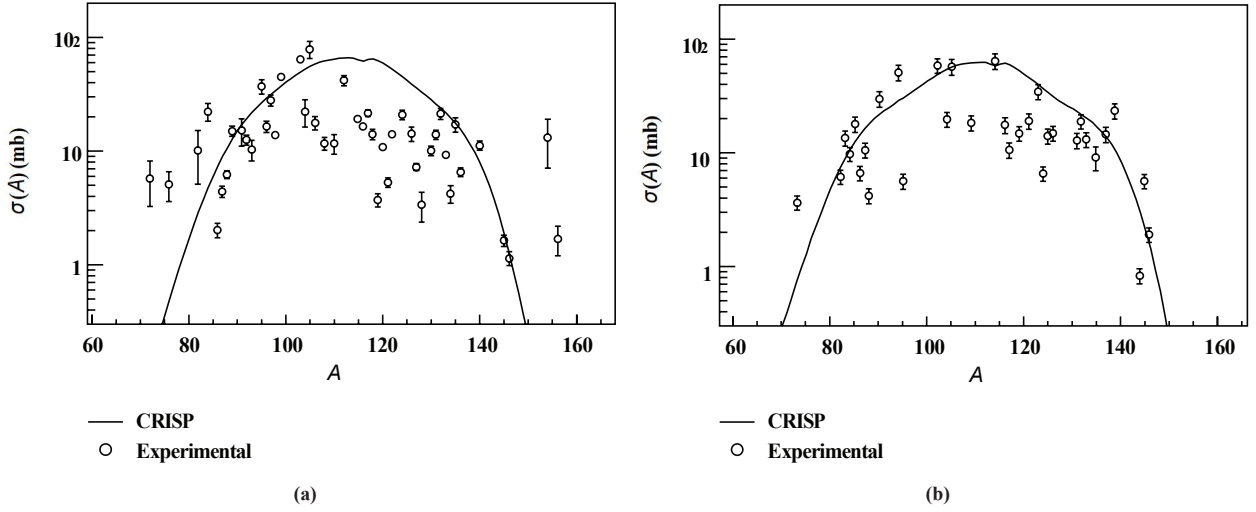

Figure 1: Cross section of fragments produced in the fission of $^{241}$Am (a) and $^{237}$Np as a function of fragment mass (b) induced by 660 MeV protons. Values from the systematic study of [9] for the parameters corresponding to the three fission modes considered in this work (—) were used. Open symbols (○) are the experimental data [10].

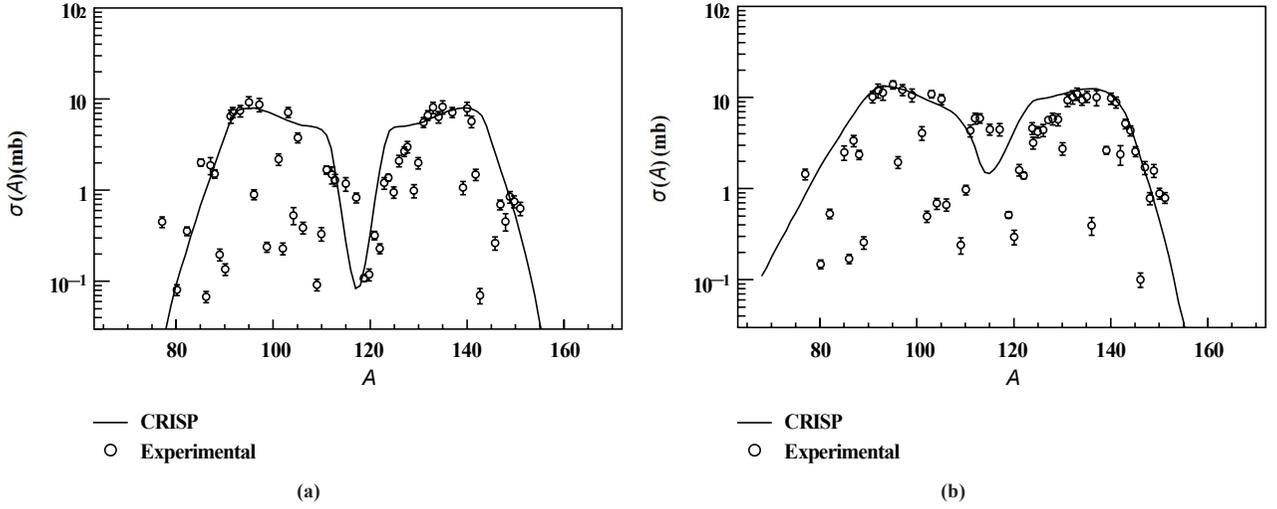

Figure 2: Cross sections for fragments produced in the fission of $^{238}$U induced by bremsstrahlung of 50 MeV (a) and 3500 MeV (b) endpoint energies as a function of fragment mass. Open circles (○) are experimental results [11].

the contributions of each channel, $\sigma_i(A, Z)$, which are usually written in the form

$$\sigma(A,Z) = \left\{ \sum_{i=1,2} \left[ \frac{K_i^L}{\sqrt{2\pi}\Gamma_i^L} \exp\left(-\frac{(A-A_i^L)^2}{2(\Gamma_i^L)^2}\right) \right. \right.$$
$$\left. + \frac{K_i^H}{\sqrt{2\pi}\Gamma_i^H} \exp\left(-\frac{(A-A_i^H)^2}{2(\Gamma_i^H)^2}\right) \right]$$
$$\left. + \frac{K_S}{\sqrt{2\pi}\Gamma_S} \exp\left(-\frac{(A-A_S)^2}{2(\Gamma_S)^2}\right) \right\} \quad (12)$$
$$\times \frac{1}{\sqrt{2\pi}\Gamma_Z} \exp\left(-\frac{(Z-Z_f)^2}{2\Gamma_Z^2}\right),$$

where the summation runs over the asymmetric modes. The parameters for the symmetric mode are $K_S$, $A_S$, and $\Gamma_S$, while $K_i^{H(L)}$, $A_i^{H(L)}$, and $\Gamma_i^{H(L)}$ are the parameters for the heavy (light) fragment produced in the asymmetric mode $i$. For the atomic number distribution the parametrization used is

$$Z_f = \mu_1 + \mu_2 A, \quad (13)$$

for the most probable atomic number of the fragment, and

$$\Gamma_Z = \nu_1 + \nu_2 A, \quad (14)$$

for the width of the atomic number distribution. $\mu_1$, $\mu_2$, $\nu_1$ and $\nu_2$ are fitting parameters.

It is important to stress that the evaporation process has as input the distribution of nuclei obtained at the end of the intranuclear cascade, therefore the fissioning nucleus may be diferent from the cascade residual nucleus because some neutrons, protons, and/or alpha-particles are allowed to evaporate before fission takes place. The efects of mass and energy distribution of the fissioning nucleus have been discussed elsewhere [40].

It is important to notice that all the parameters used in the CRISP code are fitted to a large number of experimental data for many diferent nuclei in diferent energy ranges. Also the probes can be as diferent as protons, tagged photons, Bremsstrahlung photons, or electrons. In this way, the method is reliable because it is not an *ad hoc* fitting to specific nucleus, reaction, or energy.

## 5. Results and Discussions

We used the CRISP code to calculate cross sections for the formation of fragments in nuclear reactions. As explained above, the main sources of fragments in the reactions studied here are spallation and fission processes, and in this section we show the cross sections for fission fragments and for spallation products.

*5.1. Fission Reactions.* The relevant parameters in Dostrovsky's empirical formulas (see (6)) for evaporation process are shown in Table 2. This set of parameters were used for all reactions studied in this work. The parameters for fission fragments calculations in formula (12) are shown in Table 3. For $^{241}$Am, $^{237}$Np, and $^{238}$U cases, the values for width and position (Table 3, fisrt column) of the fission modes were taken from a low-energy systematics [9]. The relative intensities of each fission mode for $^{241}$Am and $^{237}$Np were considered fixed, with their values given in Table 3. In the case of $^{238}$U the relative intensities were calculated according to gaussian expressions depending on the fissioning system mass numbers.

There is not any systematic study of the multimodal parameters in the mass region of Pb. To obtain those parameters, we used a sigmoid fit for $K_S$ and gaussian fits for $K_1$ and $K_2$, all depending on the mass of the fissioning system. For heavy fragment distributions the peak for the asymmetric modes are obtained by:

$$A_1^H = b_1 A_f + b_2,$$
$$A_2^H = b_3 A_f + b_4, \qquad (15)$$

while the width for each mode is assumed to be constant:

$$\Gamma_S = b_5,$$
$$\Gamma_1^H = b_6, \qquad (16)$$
$$\Gamma_2^H = b_7.$$

The values for each of these parameters corresponding to the best fit are presented in Table 4.

In Figure 1 we show results obtained for proton-induced fission. We observe that the results obtained with the CRISP code give a good description of the experimental data, although the data are spread over a large range.

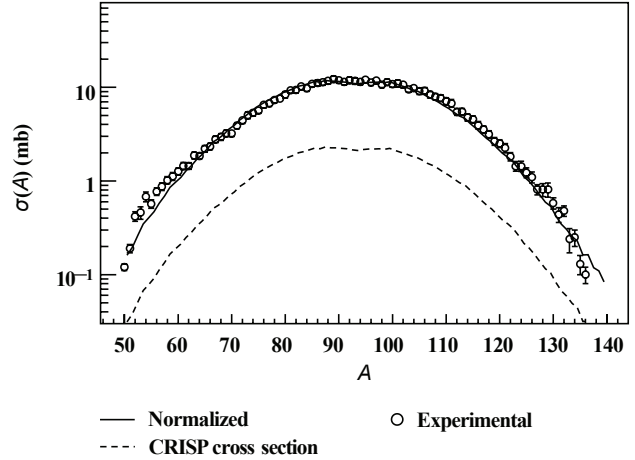

Figure 3: Cross sections for fragments produced in the fission of $^{208}$Pb induced by 500 MeV protons as a function of mass.

Since CRISP can be used also for photon-induced reactions, we calculated the mass distributions of fragments for fission induced in $^{238}$U by Bremsstrahlung photons with endpoint energies at 50 MeV and 3500 MeV. The results are presented in Figure 2. Also in this case the data do not show good resolution, but we can observe that the results with CRISP give a good general description of the experimental data.

A more precise experiment was performed for proton-induced fission on $^{208}$Pb. In this case the experimental results are very precise. In Figure 3 we show the experimental data compared to the calculation results. We observe a very good agreement between experiment and calculation with the CRISP code, especially about the overall form of the distribution. Such an agreement is better attested when the result is normalized to the data.

For all results shown in Figures 1, 2, and 3, the evaporation of fragments were performed. This was achieved by using the already presented expression (12) to obtain the mass and atomic numbers of heavy and light fragments. By first approximation, the calculated excitation energy of the fissioning system was divided between the fragments which continued the evaporation process till the stopping criterion is reached again, as explained in Section 3.

*5.2. Spallation Reactions.* Nuclei with mass number $A < 220$ present low fission probabilities. In these cases the most probable reaction channel is the spallation, when the evaporation continues till the residual nucleus is cold, without fissioning. The nucleus at the end of the evaporation is then called spallation product.

With the CRISP code, using the parameters for evaporation and fission competition as described above, we can calculate also spallation product distributions. In Figure 4 we show our results and compare them to experimental data. In general we observe a fair agreement with data, since the well-known spallation parabola as calculated by us show position and width in good agreement with experimental results. For the most probable products, also the absolute cross sections, are in good agreement with data, although for some spallation products the agreement is not good. Similar behaviour occurs for the case of spallation on $^{147}$Au, shown in Figure 5.

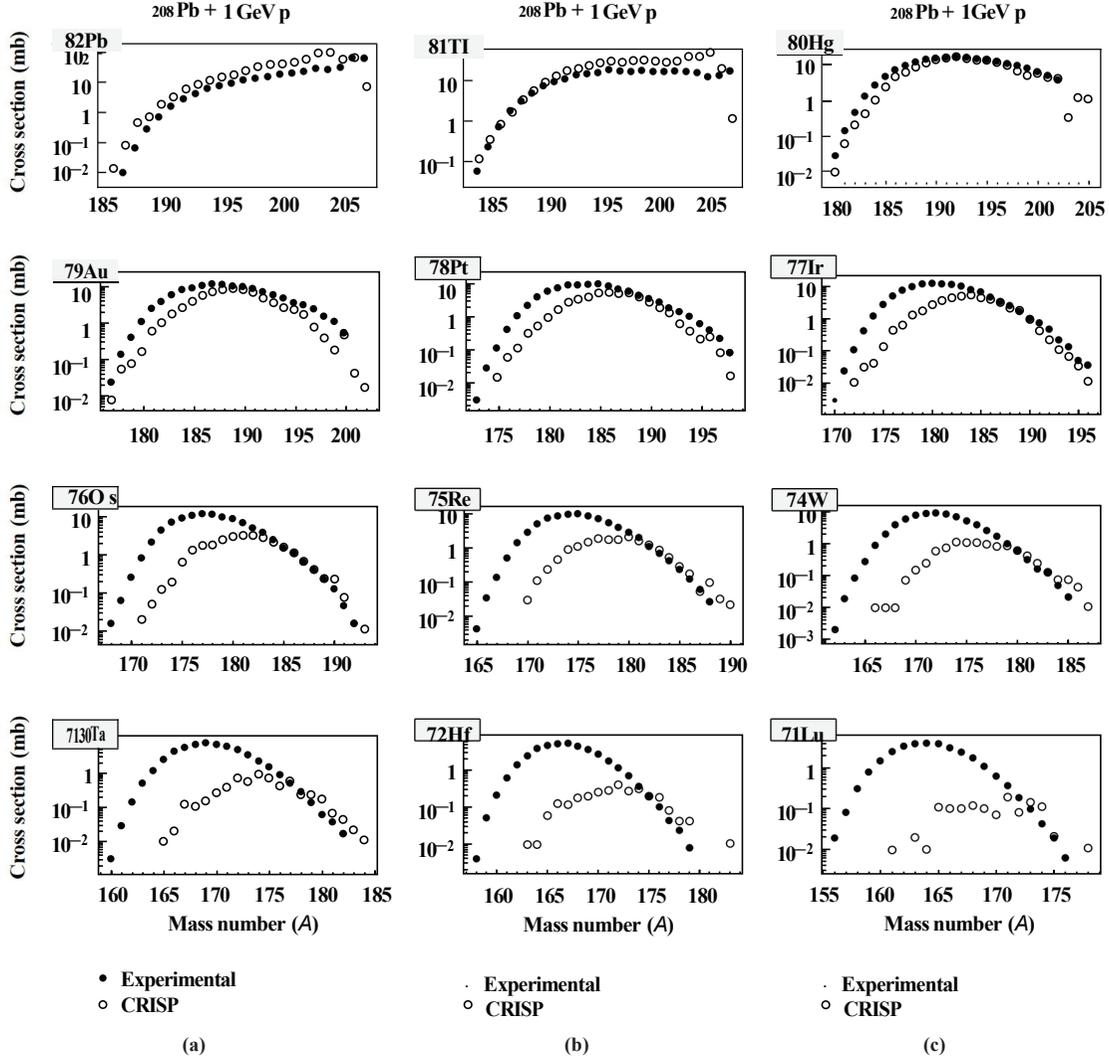

Figure 4: Cross section for fragments produced in spallation of $^{208}$Pb induced by 1 GeV protons.

In fact, these results show that it is extremely difficult to obtain good results simultaneously for several different reactions on a wide range of target mass numbers and for quite different energies. The main difficulty comes from the fact that, at the end of the intranuclear cascade, we have in general excited residual nuclei that may be far from the line of stability. Since most of the nuclear models and their parameters are determined for cold stable nuclei, we do not have precise information on the structure of all nuclei that are formed during nuclear reactions at intermediate and high energies. Improvements on mass formula and on shell effects can lead to better agreement between calculations and experiments.

## 6. Conclusion

In this work we addressed the problem of fragment production in nuclear reactions. This is a relevant issue in the development of ADS, as it is directly related to the study of damage induced in materials used in ADS.

We discussed the importance of reliability of calculation, and show that the model implemented in the CRISP code is developed with reduced number of free parameters and care in reproducing accurately the physical process that occurs during the nuclear reaction. The parameters appearing in the mass formula were obtained by fittings to the nuclear mass data. The parameters for evaporation and fission were fitted simultaneously to several results for fission and spallation cross sections on several nuclei and with different probes at many different energies. For these reasons, the code can be used to calculate several observables in different reactions induced by probes like photons, electrons, protons, and neutrons in a large energy range on nuclei with mass from $A \approx 12$ to $A \approx 240$.

We use the CRISP code for calculating the production of nuclide in nuclear reactions induced by intermediate- or high-energy probes. Proton and photon reactions on actinide and preactinide nuclei were considered. The most important mechanisms of fragments production, spallation, and fission were studied in details. We show that the results are in qualitative good agreement with experimental data available.

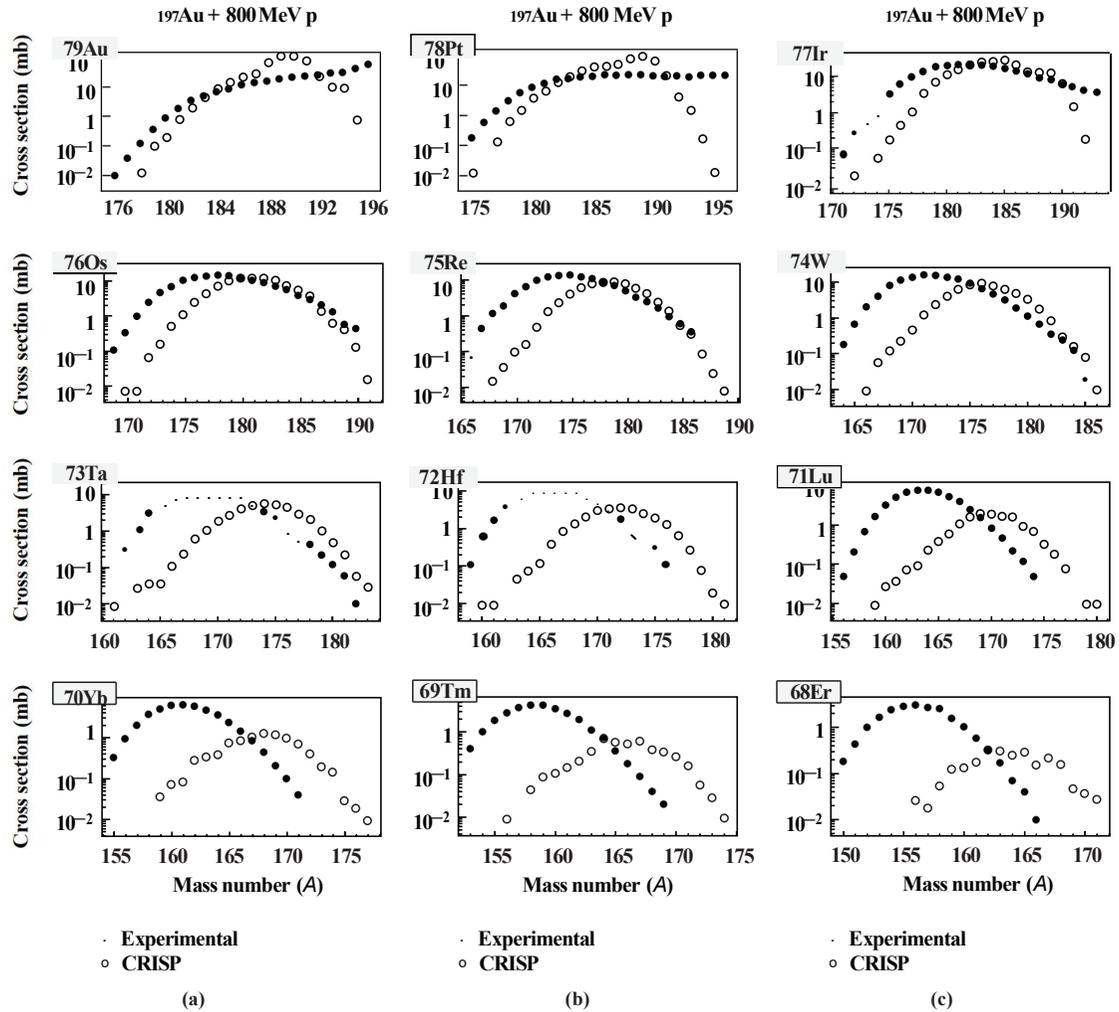

Figure 5: Cross section for fragments produced in spallation of $^{197}$Au induced by 800 MeV protons.